\documentclass[journal=jacsat,manuscript=article]{achemso}
\usepackage[version=3]{mhchem} 

\usepackage{graphicx}

\usepackage{epstopdf}

\author{O. D. D. Couto Jr.}
\email{odilon@ifi.unicamp.br}
\altaffiliation{Current address: Instituto de F\'{i}sica "Gleb Wataghin", Universidade Estadual de Campinas, 13083-859, Campinas, S\~{a}o Paulo, Brazil}
\affiliation[University of Sheffield]
{Department of Physics and Astronomy, University of Sheffield, Sheffield S3 7RH, UK}
\author{D. Sercombe}
\email{php10ds@sheffield.ac.uk}
\affiliation[University of Sheffield]
{Department of Physics and Astronomy, University of Sheffield, Sheffield S3 7RH, UK}
\author{J. Puebla}
\affiliation[University of Sheffield]
{Department of Physics and Astronomy, University of Sheffield, Sheffield S3 7RH, UK}
\author{L. Otubo}
\affiliation[IPEN]{Instituto de Pesquisas Energ\'{e}ticas e Nucleares - IPEN, S\~{a}o Paulo, Brazil}
\author{I. J. Luxmoore}
\author{M. Sich}
\author{T. J. Elliott}
\author{E. A. Chekhovich}
\author{L. R. Wilson}
\author{M. S. Skolnick}
\affiliation[University of Sheffield]
{Department of Physics and Astronomy, University of Sheffield, Sheffield S3 7RH, UK}
\author{H. Y. Liu}
\affiliation[University College London]
{Department of Electronic and Electrical Engineering, University College London, London WC1E 7JE, UK}
\author{A. I. Tartakovskii}
\email{a.tartakovskii@sheffield.ac.uk}
\affiliation[University of Sheffield]
{Department of Physics and Astronomy, University of Sheffield, Sheffield S3 7RH, UK}

\title[\texttt{achemso} demonstration]
{Effect of the GaAsP shell on optical properties of self-catalyzed GaAs nanowires grown on silicon}

\begin{document}
\begin{abstract}
We realize growth of self-catalyzed core-shell GaAs/GaAsP nanowires (NWs) on Si substrates using molecular-beam epitaxy. Transmission electron microscopy (TEM) of single GaAs/GaAsP NWs confirms their high crystal quality and shows domination of the zinc-blende phase. This is further confirmed in optics of single NWs, studied using cw and time-resolved photoluminescence (PL). A detailed comparison with uncapped GaAs NWs emphasizes the effect of the GaAsP capping in suppressing the non-radiative surface states: significant PL enhancement in the core-shell structures exceeding 2000 times at 10K is observed; in uncapped NWs PL is quenched at 60K whereas single core-shell GaAs/GaAsP NWs exhibit bright emission even at room temperature. From analysis of the PL temperature dependence in both types of NW we are able to determine the main carrier escape mechanisms leading to the PL quench.
\end{abstract}


\section{Introduction}
\label{introduction}

Incorporating light-emitting components into Si microelectronics has been the driving factor behind the development of Si photonics for the last twenty years~\cite{Yan09}. The difficulty of such implementation stems from the lattice mismatch between Si and III-V semiconductors commonly used for generation of light. Transfer of III-V structures onto Si substrates is also driven by significant reduction in substrate costs, and will potentially facilitate large scale production of highly efficient III-V photo-voltaic elements. Semiconductor nanowires (NWs) present highly versatile nanostructures less reliant on the lattice-matching with the substrate than bulk semiconductors~\cite{Joyce11}. Successful growth of nearly lattice-matched GaP NWs on Si was initially demonstrated using metal-organic vapor-phase epitaxy and gold nano-particles as a seed ~\cite{Martensson04}. Recently, gold contamination problems have been overcome by successful growth of catalyst-free III-V NWs on Si using molecular beam epitaxy (MBE)~\cite{Colombo08,Cirlin10,Plissard10}. Despite many reports on their high crystal quality, few reports on the optical properties of MBE grown GaAs structures exist~\cite{Paek10}. Optical measurements on single self-catalyzed NWs on Si would be important for in-depth understanding and design of electronic properties of NWs, and for future incorporation of quantum dot nano-structures in III-V light-emitting devices monolithically grown on Si~\cite{Chen11,Yeo12}.

In general, improvement in the optical properties of NWs is achieved by capping of the NW core with a higher band gap material, suppressing  non-radiative carrier escape via the surface states \cite{Perera08,Demichel10,Montazeri10}. The effects of AlGaAs passivation on the optical properties of GaAs NWs have been reported \cite{Perera08,Demichel10}. In contrast, GaAsP-passivated MBE-grown NWs have not yet been demonstrated  despite the fact that Ga(As)P-capped NWs may potentially be more attractive as such material will naturally have low density of surface states, and will be less prone to oxidation compared to AlGaAs.

In this letter, we realize catalyst-free MBE-grown GaAs/GaAsP core-shell NWs on Si(111) substrates (with nominally 15$\%$ phosphorus). TEM studies show high crystal quality of the capped NWs and demonstrate that the NWs mostly appear in the zinc-blende (ZB) phase. We demonstrate the growth of low NW density samples, which enables on-chip optical addressing of isolated NWs on unpatterned Si substrates. In this way, by using continuous-wave (CW) and time-resolved (TR) micro-photoluminescence ($\mu$-PL) spectroscopy, we investigate optical properties of single NWs. We find that GaAsP capping leads to greatly enhanced PL compared with uncapped GaAs, with single NW PL readily observed at room temperature. Measurements of PL temperature dependence in both types of NW enable identification of the main carrier escape mechanisms and determination of the activation energies corresponding to the characteristic non-radiative processes. Our results show the detrimental effect of surface states and oxidation in uncapped GaAs structures, which is otherwise strongly suppressed by GaAsP capping in GaAs/GaAsP NWs.  The capping also leads to the PL blue-shift with respect to both ZB and wurzite (WZ) GaAs band-gaps, characteristic of strained core material, occurring despite the presence of a large free surface of the nanostructures and relatively low concentration of phosphorus.  At low temperatures, carrier lifetimes $\approx$1~ns  and relatively large inhomogeneous PL width of the order of 20-50 meV measured in single NWs indicate carrier localization in quantum-dot-like nanoscale potential minima possibly caused by strain. On the contrary, uncapped GaAs NWs exhibit a wide range of carrier lifetimes from $<0.3$ up to $\approx$~7~ns owing to a mixture of type-I and type-II electron-hole alignment in a single NW.

\section{Experimental procedure}

\subsection{Growth}

\begin{figure}[t]
\includegraphics[width=0.6\textwidth]{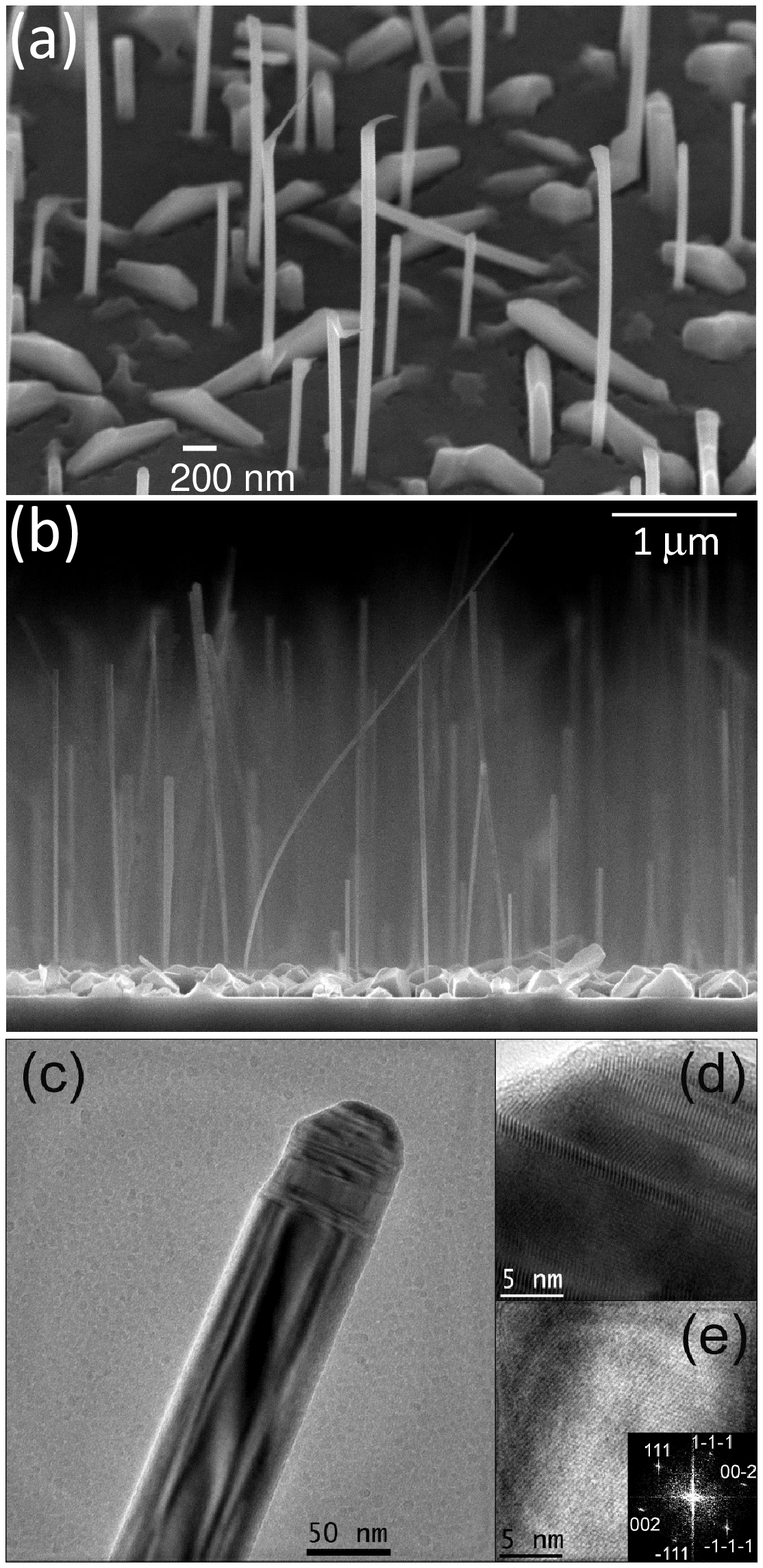}
   \caption{\label{Fig1} High-resolution scanning electron microscopy (SEM) image of (a) core-shell GaAs/GaAsP nanowires and (b) uncapped GaAs nanowires. Low magnification TEM image (c), high-resolution TEM images of the tip (d), and of the middle part (e) of a GaAs/GaAsP NW. The inset shows the FFT of image (e).}
\end{figure}

NWs were grown on p-type Si(111) substrates using a Veeco Gen930 MBE equipped with standard effusion cells for group III elements and Veeco valve cracker cells supplying arsenic (As4) and phosphorus (P2). A pyrometer was used to measure the substrate temperature in situ, which was calibrated to the onset temperature for desorption of the (001) GaAs surface native oxide. Two structures have been studied. The capped GaAs NWs were grown at 640~$^{\circ}$C. After the GaAs growth, the sample was cooled down to 500~$^{\circ}$C and passivated with a GaAs$_{1-x}$P$_x$ cap layer with the nominal thickness of 70 nm. The nominal concentration of phosphorus  was $x$=0.15.  For the other sample, NWs were grown under very similar conditions but at a slightly lower temperature of 635~$^{\circ}$C. No passivation was carried out in this sample so that the GaAs surface was exposed to air.

\ref{Fig1} shows high-resolution scanning electron microscopy (SEM) images obtained for GaAs/GaAsP NWs in (a) and uncapped GaAs NWs in (b).  Both types of NWs are found to have lengths in the range of 1~-~3~$\mu$m. Diameters in capped and uncapped NWs range between 50-120 nm and 30-80~nm, respectively, depending on the NW position relative to the center of the wafer. The image on \ref{Fig1}(a), taken from a region at the center of the Si wafer, shows one of the highest GaAs/GaAsP NW densities found on the surface of sample of 1.5~NW/$\mu$m$^2$. A typical separation between NWs is approximately 1~$\mu$m. Still, well isolated NWs could be readily found and addressed optically. All optical studies presented below were carried out in sample regions where light emission by single NWs could be unambiguously identified. Similarly, single NWs could be identified in PL in the uncapped NWs sample, where NW density varies from approximately 2~NW/$\mu$m$^{2}$ at the center of the 3-inch Si wafer to $\leq$~0.5~NW/$\mu$m$^{2}$ at its edge, as was observed in detailed SEM analysis.

\ref{Fig1}(c), (d), and (e) show high resolution transmission electron microscopy (TEM) measurements on a single GaAs/GaAsP NW. They were performed using a JEM 2100 (200 kV) TEM microscope. Specimens were prepared by ultrasonicating the nanowires in isopropanol for 1 min followed by dispersal onto collodion coated copper grid. Figure 1(c) depicts a representative TEM image of capped NWs from the central area of the wafer. As we observe, the core thickness is about 50 nm. In general, NWs present very uniform diameters along their length with some defects only found close to their tips. Defects are shown in more detail in \ref{Fig1}(d), where it is possible to identify some stacking faults.  \ref{Fig1}(e) exhibits a high resolution image from the NW middle part. The fast Fourier transform from this image (inset) shows that most part of the NW is formed of ZB structure with a growth axis along the $[\bar{1}\bar{1}1]$ direction. TEM measurement in other NWs from the same region showed that in general NW have ZB  structure even at the tip, where stacking faults are related to twinning of the crystal planes~\cite{Zardo09}. In some NWs, however, we observed regions with WZ structure among the ZB segments~\cite{Krogstrup10}.

\subsection{Optical measurements}
\label{experiments}

The optical properties of the NWs were studied in CW and TR $\mu$-PL experiments. In CW measurements, optical excitation at 1.893~eV was performed with a diode laser focused on the sample in $\approx$1~$\mu$m spot. Owing to the low  NW surface density achieved by the growth, PL from single NWs could be collected using a high numerical aperture lens positioned above the substrate from as-grown single NWs standing on Si. We also carried out comparative optical studies on NWs removed from the substrate and deposited on a clean Si substrate, where we obtained similar results. Spectra were recorded using a single-spectrometer equipped with a charge-coupled device (CCD) camera. Samples were placed on a cold-finger cryostat which allowed for temperature control from 5 to 300~K.

TR $\mu$-PL experiments were carried out using two different setups. On the GaAs/GaAsP sample, measurements were performed in an $8$ meV spectral range using a streak camera with the time resolution below 5~ps.  The TR experiments on the uncapped GaAs NWs were performed using an avalanche photodiode (APD) with temporal resolution of approximately 350~ps in a spectral range of 1 meV. In both TR set-ups, non-resonant optical pumping was realized using a pulsed Ti-Sapphire laser producing 2~ps pulses. Laser excitation used for samples GaAs/GaAsP and GaAs was at 1.71 and 1.595~eV, respectively, at energies more than 100 meV above NW PL in both samples.

\section{Photoluminescence results}


\begin{figure}[t]
\includegraphics[width=0.6\textwidth]{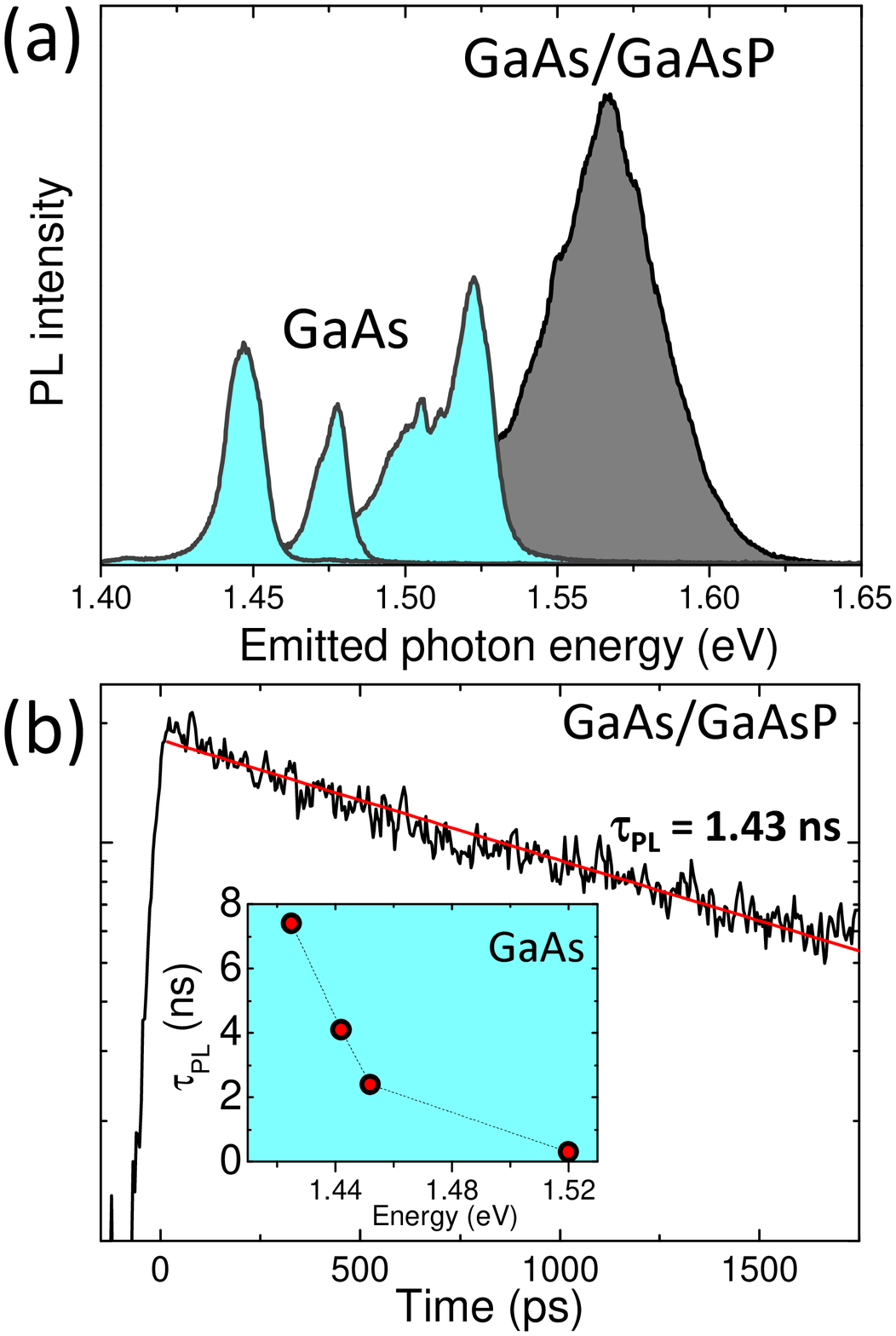}
   \caption{\label{Fig2}(a) Photoluminescence (PL) spectra measured for individual GaAs/GaAsP core-shell nanowires (dark grey) and uncapped GaAs nanowires (light grey) at T=10K. For both samples spectra are excited at 1.893~eV using a diode laser. GaAs/GaAsP (GaAs) PL is measured with a laser power $P_{\textmd{ex}} = 100$~nW ($100\mu$W) and acquisition time 5 s (10 s), i.e. PL from the capped NWs is about 2000 times stronger than from the uncapped structures. (b) Time-resolved PL measured for a single GaAs/GaAsP NW. The inset shows dependence of PL lifetime on the detection energy for single uncapped GaAs NWs.}
\end{figure}

\ref{Fig2}(a) shows typical $\mu$-PL spectra (T=10K) measured for single GaAs/GaAsP NWs (dark grey) and uncapped GaAs NWs (light blue) at 10~K. For GaAa/GaAsP NW, laser power $P_{\textmd{ex}}$ = 100~nW and acquisition time of 5~s were used. The PL peak is observed at $\approx$1.57~eV, exceeding the maximum PL energy of 1.52 eV observed for uncapped GaAs NWs and above the ZB bulk band-gap of 1.51 eV. GaAs/GaAsP NW PL peaks were observed to have linewidths between 20 and 50 meV. Very similar PL spectra are found on a length scale of a few mm, thus demonstrating the uniformity of the NWs on a large wafer region (which was also confirmed by TEM analysis).

A very different picture is observed for uncapped GaAs NWs. The 3 single NW spectra shown in \ref{Fig2}(a) were also obtained within a few mm distance along the sample surface. A marked PL peak energy variation is observed exceeding 100 meV. At the same time, individual PL peaks are relatively narrow with widths down to 5-20 meV. Also note that PL from bare NWs is much weaker: in \ref{Fig2}(a) spectra are recorded for 100 $\mu$W excitation power and acquisition time of 10~s. Thus, in the given example the PL for the core-shell NW is more than 2000 times stronger compared with the uncapped structures (in \ref{Fig2}(a), powers much lower than PL saturation were used).


In order to gain deeper insight in the optical and structural properties of NWs, TR PL experiments were carried out. For all studied capped NWs, carrier lifetimes ($\tau_{PL}$) from 1-1.5 ns were found. This is shown in \ref{Fig2}(b) for a single GaAs/GaAsP NW for which $\tau_{PL}$ = 1.43~ns. On the other hand, for uncapped structures, a wide range of $\tau_{PL}$ values was found depending on the NW emission energy. This is shown on the inset of \ref{Fig2}(b) where $\tau_{PL}$=7.4 ns for a NW emitting at 1.425 eV, but decreases as NW energy increases. For a NW emitting around the ZB band-gap of 1.52 eV, our time-resolution limit of 0.3 ns is reached. Similar results were obtained for several uncapped NWs.

\section{Discussion}

\begin{figure}[t]
\includegraphics[width=0.6\textwidth]{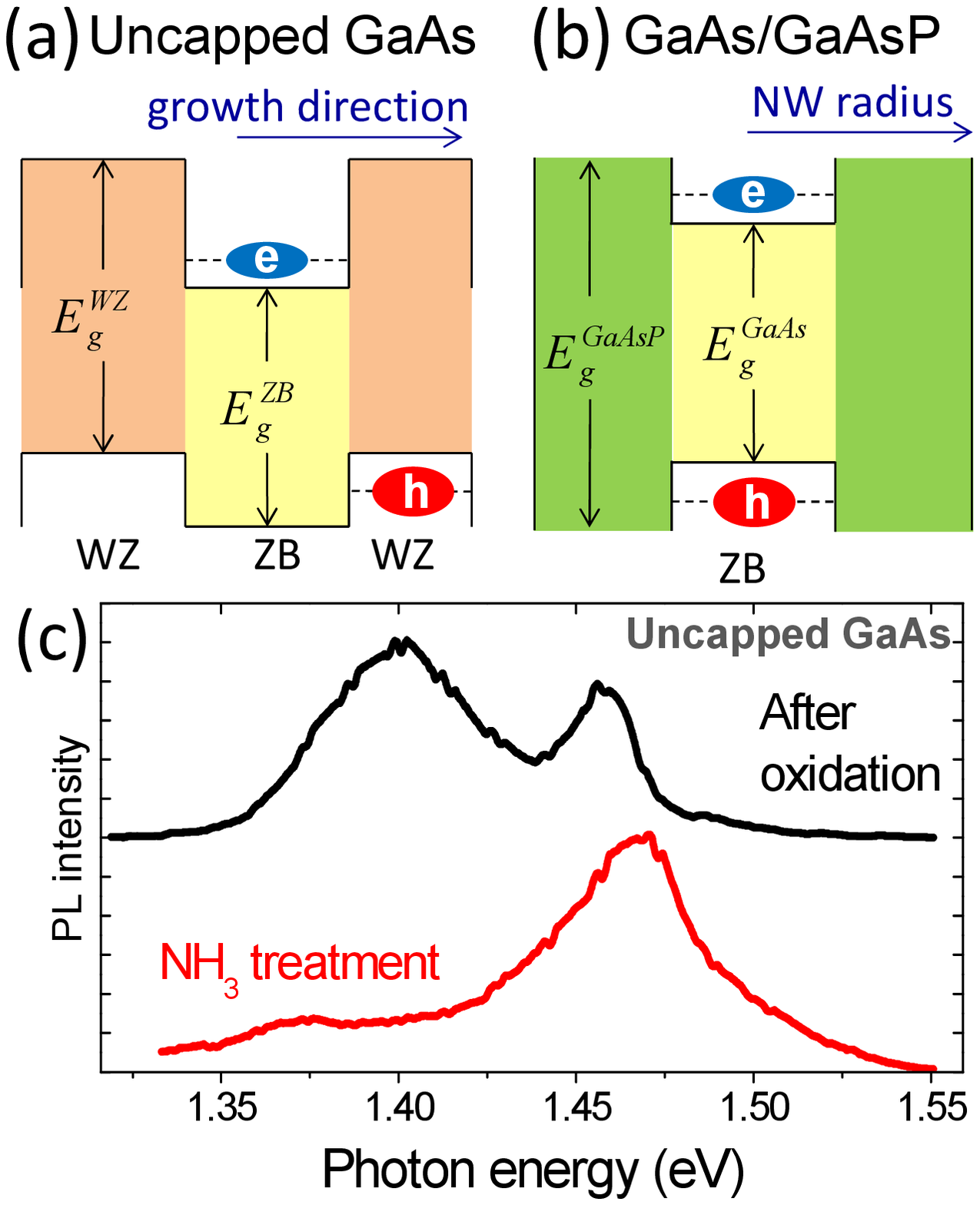}
   \caption{\label{Fig3} Schematics of band-structure of (a) an uncapped GaAs NW and (b) a GaAs/GaAsP core-shell NW. GaAs/GaAsP NW has ZB structure and electrons and holes strongly confined withing the GaAs core (the band-structure along the radial direction in the wire is shown), whereas uncapped GaAs NW shows a mixture of zinc-blende (ZB) and wurzite (WZ) structure (the band-structure along the wire axis is shown). (c) PL spectra measured on single uncapped GaAs NWs: (i) after exposure to air (black curve) showing strong peak at around 1.4 eV and weak NW PL at 1.46 eV and (ii) after exposure to air and washing in NH$_3$ (red curve), which almost fully restores PL of the NWs.}
\end{figure}

The PL measurements in \ref{Fig2} allow the following conclusions to be made about the band structure of the two types of NWs, which is schematically depicted in \ref{Fig3}~(a) and (b). The emission energy range for uncapped NWs spans from approximately 1.40 to 1.52~eV. Such broad emission range reflects the type-II band alignment along the NW growth direction [illustrated in \ref{Fig3}(a)] originating from segments of the ZB and WZ crystal structure in III-V NWs~\cite{Murayama94,Bao08,Spirkoska09,Pemasiri09,Akopian10,Zhang10}. In such circumstances, electrons are confined in the conduction band minimum of ZB structure (with the bandgap $E^{ZB}_g$ = 1.515~eV) while holes are confined at the maximum of the valence band of the adjacent WZ segment with bandgap $E^{WZ}_g$ measured between 1.5 and 1.54~eV~\cite{Heiss10,Spirkoska09,Hoang09}. The wide range of emission energies observed in our studies arises from variation of the thicknesses of the WZ and ZB segments, leading to variation of the carrier confinement. The lowest PL energy around 1.405~eV observed in \ref{Fig2}(a) reflects a valence band offset between WZ and ZB structures of approximately 115~meV in good agreement with calculated values~\cite{Murayama94,Heiss10}, and possibly arises from e-h recombination at the boundary between large ZB and WZ segments.

The conclusions about the complex ZB/WZ structure in uncapped NWs are supported by the energy dependence of the carrier lifetime $\tau_{PL}$, shown on the inset of \ref{Fig2}(b). Indeed, the long $\tau_{PL}$ up to 7.4 ns are characteristic of indirect exciton recombination \cite{Spirkoska09,Pemasiri09}. The shortening of $\tau_{PL}$ with increasing energy (arising due to the reduced size of the ZB or WZ segments) reflects stronger overlap between the electron and hole, as the carrier confinement within the ZB and WZ segments becomes weaker. Additional lifetime shortening may be related to contribution of carrier escape to non-radiative centres on the NW surface. Around the energy of 1.52 eV, where the life-time becomes shorter than 0.3 ns, both fast type-I exciton recombination and non-radiative processes at the NW surface will dominate, thus leading to the PL signal decay beyond our resolution.

\ref{Fig3}(b) shows schematically the type-I band alignment along the NW radius of the core-shell structures, which could be deduced from the optical studies discussed above. The higher emission energy compared to ZB and WZ GaAs bandgaps implies that these NWs are strained, as a consequence of the lattice-mismatched GaAs core and GaAsP shell~\cite{Gourley84,Montazeri10}. Relatively large PL linewidths are related to the (radial) confinement potential variations which can appear due to strain inhomogeneities or the twining effect of the ZB planes~\cite{Zardo09}. Carrier lifetimes $\tau_{PL}$ in the range of 1-1.5 ns are similar to lifetimes in type-I self-assembled QDs where excitons are typically localized on the length scales smaller than the exciton Bohr radii in corresponding bulk materials. This implies that PL in GaAsP-capped NWs is likely to originate from strongly confined excitons. Besides this, the PL temperature dependence we show below demonstrates that excitons are strongly confined in the core part of the NW and are well-isolated from the surface. Note, efficient carrier confinement in the GaAs core is further confirmed by the fact that no PL corresponding to the GaAsP shell was detected.

\section{Temperature-dependent PL measurements and role of the surface states in PL quenching}

\ref{Fig3}(c) presents evidence that suppression of PL in uncapped structures is due to the detrimental effect of the surface states. As shown in \ref{Fig2}(a), PL is significantly weaker in uncapped NWs as compared to core-shell structures. However, we also observed that the emission properties from the uncapped NWs strongly degraded with time (sometimes on a time-scale of a few hours), especially after exposure to air. In such degraded NWs, in addition to a reduction of initial PL intensity, we observed the appearance of a broader emission band between 70 and 130~meV below the NW emission energy. This band possibly corresponds to the oxygen impurities, most probably arising due to the oxidation of the NW surface~\cite{Skowronski90}. As shown in \ref{Fig3}(c), original PL characteristics can be restored by washing NWs in ammonia (NH$_3$)~\cite{noteNH3}. No such degradation was detected in GaAs/GaAsP core-shell structures.

\begin{figure}[t]
\includegraphics[width=0.6\textwidth]{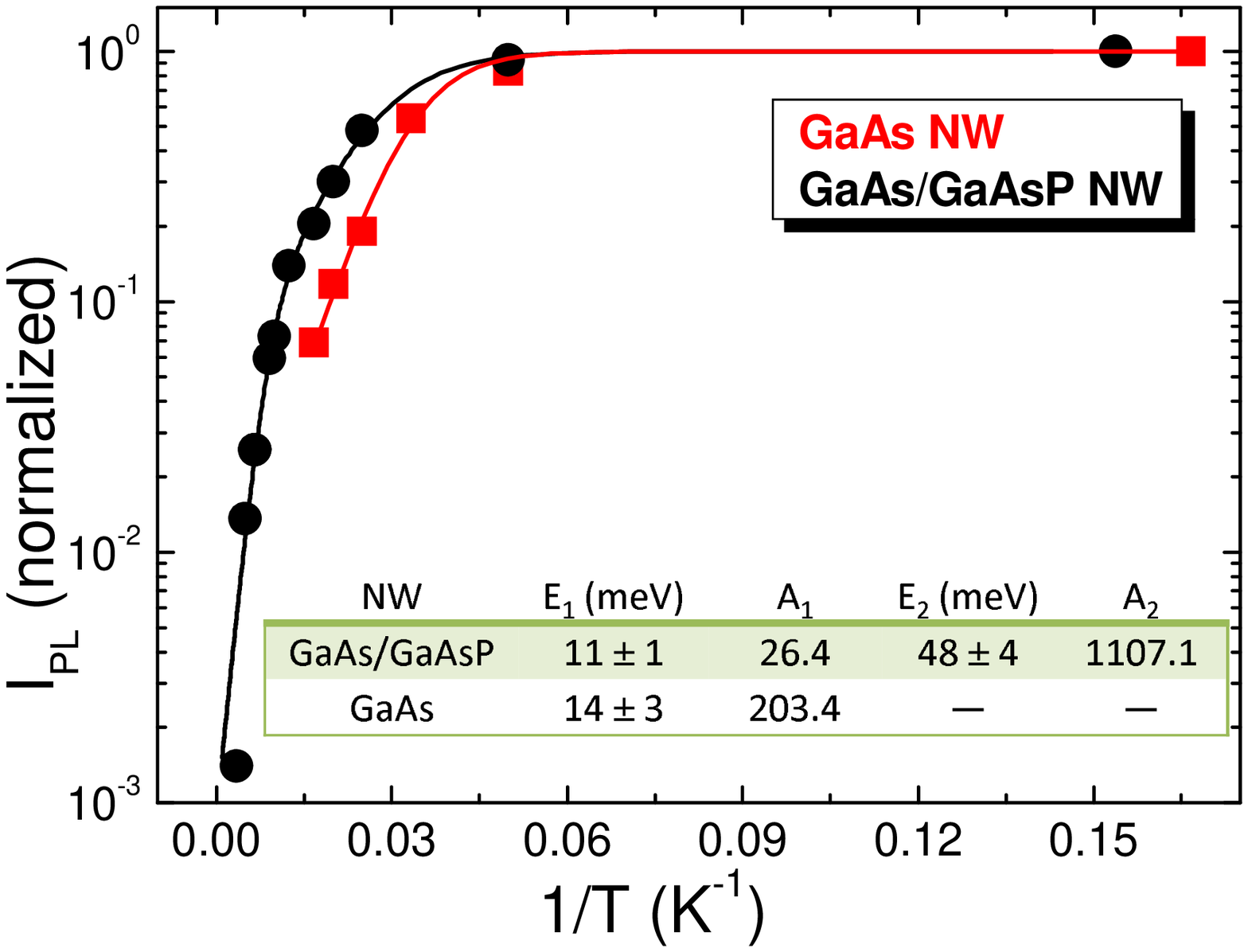}
\caption{\label{Fig4} PL intensity for a core-shell GaAs/GaAsP NW (circles) and an uncapped GaAs NW (squares) as a function of temperature. Both PL intensities are normalized by their values measured at $T\approx 10$K. Activation energies and non radiative recombination ratios obtained from fitting with Eq.1 are summarized on the inset.}
\end{figure}

Further insight into the electronic structure of the NWs and the role of the surface states was gained from PL temperature dependence. \ref{Fig4} shows the temperature dependence of the normalized integrated PL intensity for two single NWs: a GaAs/GaAsP NW (circles) with central emission peak at 1.61 eV (at 6~K) and an unpassivated GaAs (squares) emitting at 1.475 eV (at 6~K). As we can observe, PL decays relatively faster with T for the case of the uncapped structure. Above 60 K the noise-to-signal ratio becomes close to one and PL cannot be detected. In contrast, the GaAs/GaAsP NW PL decays much slower with T and can be easily detected at room temperature.

For the GaAs/GaAsP structure, the experimental data can be well described in the whole temperature range by a dual activation energy equation~\cite{Bimberg71},

\begin{equation}
\frac{I_{PL}(T)}{I_0} = \frac{1}{1+ A_1~\textrm{exp}(-E_1/K_BT)+ A_2~\textrm{exp}(-E_2/K_BT)}
\label{eq1}
\end{equation}
where $I_0$ is the normalization constant and $E_1$ and $E_2$ are the activation energies describing non-radiative processes. The fitting constants $A_1$ and $A_2$ characterize the efficiencies of non-radiative processes related to $E_1$ and $E_2$. $K_B$ is the Boltzmann constant. For the uncapped structures, due the limited temperature range, we used a similar model with a single activation energy (parameters $A_1$, $E_1$). The values obtained from the fittings (solid lines) for the two types of NWs are displayed on the inset in \ref{Fig4}. The values of $E_1$ are similar for the two types of NWs. These are possibly associated with non-radiative recombination centres at the GaAs surface~\cite{Titova06}. Interestingly, the value of $A_1$ is approximately 8 times smaller for the passivated structure, which is a consequence of the lower density of surface states at the GaAs/GaAsP interface as compared to the GaAs surface exposed to air in the uncapped NWs.

In the core-shell NWs PL is observed up to room temperature (where their PL is still brighter than from the bare GaAs NWs at 6K). PL quenching above 100K proceeds most efficiently via the activation process characterized with $E_2$=48 meV, which can be associated with the carrier escape from the GaAs core into the GaAsP barrier. The conduction (CB) and valence (VB) band confinement potentials for the (unstrained) GaAs/GaAs$_{0.85}$P$_{0.15}$ structure can be estimated from the calculations for the GaAs/GaP core-shell system performed by Montazeri~\emph{et. al.}~\cite{Montazeri10}. Assuming a linear decrease of the potential height, as the P composition is reduced to 0.15 we estimate the CB and VB band offsets as $\Delta E_{CB}~\approx$~75~meV and $\Delta E_{VB}~\approx$~45~meV, which correspond to the expected energy barriers for electrons and holes, respectively. This rather crude estimate (not taking into account strain in our NWs) allows to attribute the value of $E_2$ to the non-radiative escape of holes from the core to the shell of the GaAs/GaAsP NWs.

\section{Conclusions}

We have investigated optical properties of single GaAs/GaAsP and uncapped GaAs nanowires grown on Si substrates using catalyst-free MBE techniques. The low nanowire density obtained by this growth method allows "on-chip" analysis of the cw and time-resolved PL response of single nano-emitters without their removal from the substrate. Considerable enhancement of PL intensity is achieved by passivation of nanowires with a GaAsP layer. When compared to similar uncapped GaAs single nanowires, GaAs/GaAsP structures also show significantly more robust optical properties in a wide range of temperatures: (i) PL from capped NWs is stronger by a factor exceeding 2000 at low T; (ii) bright PL is observed in GaAs/GaAsP at room temperature, whereas it is quenched below the detector noise level in bare GaAs nanostructures. The detailed temperature-dependent studies show that at high T, PL quenching occurs via thermal excitation of holes from the confined states in GaAs to the GaAsP core. On the other hand, bright PL in capped NWs is a consequence of a considerable suppression of non-radiative processes most probably related to the surface states. The importance of surface states originating due to oxidation and leading to PL quenching is further demonstrated by studies of uncapped GaAs NWs exposed to air for a considerable period. We show that this deterioration process can be reversed by washing the samples with ammonia, which removes the oxide layer.

We find that the introduction of the GaAsP shell results in strained GaAs core, which is evidenced by the blue shift of PL above the band-gap of zinc-blende GaAs and also relatively broad distribution of PL emission in single NWs. Recombination lifetimes from such strained core-shell structures were of the order of 1~ns at low temperatures, comparable with a high crystalline quality quantum dot structures. The reported optical and structural TEM studies lead to the conclusion that GaAs/GaAsP NWs present a high quality material with robust optical properties highly suitable for applications in nanoscale light-emitting devices (nano-lasers) and for incorporation of quantum dot nanostructures having lower band-gap than GaAs.




\acknowledgement

This work has been supported by the EPSRC Programme Grants EP/G001642/1 and EP/J007544/1, the Royal Society and ITN Spin-Optronics. J. Puebla has been supported by a CONACYT-Mexico Doctoral Scholarship.



\bibliography{bibliography}

\providecommand*\mcitethebibliography{\thebibliography}
\csname @ifundefined\endcsname{endmcitethebibliography}
  {\let\endmcitethebibliography\endthebibliography}{}
\begin{mcitethebibliography}{28}
\providecommand*\natexlab[1]{#1}
\providecommand*\mciteSetBstSublistMode[1]{}
\providecommand*\mciteSetBstMaxWidthForm[2]{}
\providecommand*\mciteBstWouldAddEndPuncttrue
  {\def\EndOfBibitem{\unskip.}}
\providecommand*\mciteBstWouldAddEndPunctfalse
  {\let\EndOfBibitem\relax}
\providecommand*\mciteSetBstMidEndSepPunct[3]{}
\providecommand*\mciteSetBstSublistLabelBeginEnd[3]{}
\providecommand*\EndOfBibitem{}
\mciteSetBstSublistMode{f}
\mciteSetBstMaxWidthForm{subitem}{(\alph{mcitesubitemcount})}
\mciteSetBstSublistLabelBeginEnd
  {\mcitemaxwidthsubitemform\space}
  {\relax}
  {\relax}

\bibitem[Yan et~al.(2009)Yan, Gargas, and Yang]{Yan09}
Yan,~R.~X.; Gargas,~D.; Yang,~P.~D. \emph{Nature Photon.} \textbf{2009},
  \emph{3}, 569\relax
\mciteBstWouldAddEndPuncttrue
\mciteSetBstMidEndSepPunct{\mcitedefaultmidpunct}
{\mcitedefaultendpunct}{\mcitedefaultseppunct}\relax
\EndOfBibitem
\bibitem[Joyce et~al.(2011)Joyce, Gao, Tan, Jagadish, Kim, Zou, Smith, Jackson,
  Yarrison-Rice, Parkinson, and Johnston]{Joyce11}
Joyce,~H.~J.; Gao,~Q.; Tan,~H.~H.; Jagadish,~C.; Kim,~Y.; Zou,~J.;
  Smith,~L.~M.; Jackson,~H.~E.; Yarrison-Rice,~J.~M.; Parkinson,~P.;
  Johnston,~M.~B. \emph{Prog. Quantum Electron.} \textbf{2011}, \emph{35},
  23\relax
\mciteBstWouldAddEndPuncttrue
\mciteSetBstMidEndSepPunct{\mcitedefaultmidpunct}
{\mcitedefaultendpunct}{\mcitedefaultseppunct}\relax
\EndOfBibitem
\bibitem[Martensson et~al.(2004)Martensson, Patrik, Svensson, Wacaser, Larsson,
  Seifert, Deppert, Gustafsson, Wallenberg, and Samuelson]{Martensson04}
Martensson,~T.; Patrik,~C.; Svensson,~T.; Wacaser,~B.~A.; Larsson,~M.~W.;
  Seifert,~W.; Deppert,~K.; Gustafsson,~A.; Wallenberg,~L.~R.; Samuelson,~L.
  \emph{Nano Lett.} \textbf{2004}, \emph{4}, 1987\relax
\mciteBstWouldAddEndPuncttrue
\mciteSetBstMidEndSepPunct{\mcitedefaultmidpunct}
{\mcitedefaultendpunct}{\mcitedefaultseppunct}\relax
\EndOfBibitem
\bibitem[Colombo et~al.(2008)Colombo, Spirkoska, Frimmer, Abstreiter, and
  i~Morral]{Colombo08}
Colombo,~C.; Spirkoska,~D.; Frimmer,~M.; Abstreiter,~G.; i~Morral,~A.~F.
  \emph{Phys. Rev. B} \textbf{2008}, \emph{77}, 155326\relax
\mciteBstWouldAddEndPuncttrue
\mciteSetBstMidEndSepPunct{\mcitedefaultmidpunct}
{\mcitedefaultendpunct}{\mcitedefaultseppunct}\relax
\EndOfBibitem
\bibitem[Cirlin et~al.(2010)Cirlin, Dubrovskii, Samsonenko, Bouravleuv, Durose,
  Proskuryakov, Mendes, Bowen, Kaliteevski, Abram, and Zeze]{Cirlin10}
Cirlin,~G.~E.; Dubrovskii,~V.~G.; Samsonenko,~Y.~B.; Bouravleuv,~A.~D.;
  Durose,~K.; Proskuryakov,~Y.~Y.; Mendes,~B.; Bowen,~L.; Kaliteevski,~M.~A.;
  Abram,~R.~A.; Zeze,~D. \emph{Phys. Rev. B} \textbf{2010}, \emph{82},
  035302\relax
\mciteBstWouldAddEndPuncttrue
\mciteSetBstMidEndSepPunct{\mcitedefaultmidpunct}
{\mcitedefaultendpunct}{\mcitedefaultseppunct}\relax
\EndOfBibitem
\bibitem[Plissard et~al.(2010)Plissard, Dick, Larrieu, Godey, Addad, Wallart,
  and Caroff]{Plissard10}
Plissard,~S.; Dick,~K.~A.; Larrieu,~G.; Godey,~S.; Addad,~A.; Wallart,~X.;
  Caroff,~P. \emph{Nanotehcnoloy} \textbf{2010}, \emph{21}, 385602\relax
\mciteBstWouldAddEndPuncttrue
\mciteSetBstMidEndSepPunct{\mcitedefaultmidpunct}
{\mcitedefaultendpunct}{\mcitedefaultseppunct}\relax
\EndOfBibitem
\bibitem[Paek et~al.(2010)Paek, Nishiwaki, Yamaguchi, and Sawaki]{Paek10}
Paek,~J.~H.; Nishiwaki,~T.; Yamaguchi,~M.; Sawaki,~N. \emph{Physica E}
  \textbf{2010}, \emph{42}, 2722\relax
\mciteBstWouldAddEndPuncttrue
\mciteSetBstMidEndSepPunct{\mcitedefaultmidpunct}
{\mcitedefaultendpunct}{\mcitedefaultseppunct}\relax
\EndOfBibitem
\bibitem[Chen et~al.(2011)Chen, Tran, Ng, Ko, Chuang, Sedgwick, and
  Hasnain]{Chen11}
Chen,~R.; Tran,~T. T.~D.; Ng,~K.~W.; Ko,~W.~S.; Chuang,~L.~C.; Sedgwick,~F.~G.;
  Hasnain,~C.~C. \emph{Nature Photon.} \textbf{2011}, \emph{5}, 170\relax
\mciteBstWouldAddEndPuncttrue
\mciteSetBstMidEndSepPunct{\mcitedefaultmidpunct}
{\mcitedefaultendpunct}{\mcitedefaultseppunct}\relax
\EndOfBibitem
\bibitem[Yeo et~al.(2012)Yeo, Malik, Munsch, Dupuy, Bleuse, Niquet, G\'{e}rard,
  Claudon, Wagner, Seidelin, Auff\`{e}ves, Poizat, and Nogues]{Yeo12}
Yeo,~I.; Malik,~N.~S.; Munsch,~M.; Dupuy,~E.; Bleuse,~J.; Niquet,~Y.;
  G\'{e}rard,~J.; Claudon,~J.; Wagner,~E.; Seidelin,~S.; Auff\`{e}ves,~A.;
  Poizat,~J.; Nogues,~G. \emph{Appl. Phys. Lett.} \textbf{2012}, \emph{99},
  233106\relax
\mciteBstWouldAddEndPuncttrue
\mciteSetBstMidEndSepPunct{\mcitedefaultmidpunct}
{\mcitedefaultendpunct}{\mcitedefaultseppunct}\relax
\EndOfBibitem
\bibitem[Perera et~al.(2008)Perera, Fickenscher, Jackson, Smith, Yarrison-Rice,
  Joyce, Gao, Tan, Jagadish, Zhang, and Zou]{Perera08}
Perera,~S.; Fickenscher,~M.~A.; Jackson,~H.~E.; Smith,~L.~M.;
  Yarrison-Rice,~J.~M.; Joyce,~H.~J.; Gao,~Q.; Tan,~H.~H.; Jagadish,~C.;
  Zhang,~X.; Zou,~J. \emph{Appl. Phys. Lett.} \textbf{2008}, \emph{93},
  053110\relax
\mciteBstWouldAddEndPuncttrue
\mciteSetBstMidEndSepPunct{\mcitedefaultmidpunct}
{\mcitedefaultendpunct}{\mcitedefaultseppunct}\relax
\EndOfBibitem
\bibitem[Demichel et~al.(2010)Demichel, Heiss, Bleusse, Mariette, and
  i~Morral]{Demichel10}
Demichel,~O.; Heiss,~M.; Bleusse,~J.; Mariette,~H.; i~Morral,~A.~F. \emph{Appl.
  Phys. Lett.} \textbf{2010}, \emph{97}, 201907\relax
\mciteBstWouldAddEndPuncttrue
\mciteSetBstMidEndSepPunct{\mcitedefaultmidpunct}
{\mcitedefaultendpunct}{\mcitedefaultseppunct}\relax
\EndOfBibitem
\bibitem[Montazeri et~al.(2010)Montazeri, Fickenscher, Smith, Jackson,
  Yarrison-Rice, Kang, Gao, Tan, Jagadish, Guo, Zou, E.Pistol, and
  Pryor]{Montazeri10}
Montazeri,~M.; Fickenscher,~M.; Smith,~L.~M.; Jackson,~H.~E.;
  Yarrison-Rice,~J.; Kang,~J.~H.; Gao,~Q.; Tan,~H.~H.; Jagadish,~C.; Guo,~Y.;
  Zou,~J.; E.Pistol,~M.; Pryor,~C.~E. \emph{Nano Lett.} \textbf{2010},
  \emph{10}, 880\relax
\mciteBstWouldAddEndPuncttrue
\mciteSetBstMidEndSepPunct{\mcitedefaultmidpunct}
{\mcitedefaultendpunct}{\mcitedefaultseppunct}\relax
\EndOfBibitem
\bibitem[Zardo et~al.(2009)Zardo, Conesa-Boj, Peiro, Morante, Arbiol, Uccelli,
  Abstreiter, and i~Morral]{Zardo09}
Zardo,~I.; Conesa-Boj,~S.; Peiro,~F.; Morante,~J.~R.; Arbiol,~J.; Uccelli,~E.;
  Abstreiter,~G.; i~Morral,~A.~F. \emph{Phys. Rev. B} \textbf{2009}, \emph{80},
  245324\relax
\mciteBstWouldAddEndPuncttrue
\mciteSetBstMidEndSepPunct{\mcitedefaultmidpunct}
{\mcitedefaultendpunct}{\mcitedefaultseppunct}\relax
\EndOfBibitem
\bibitem[Krogstrup et~al.(2010)Krogstrup, Popovitz-Biro, Johnson, Madsen,
  Nyg{\aa}rd, and Shtrikman]{Krogstrup10}
Krogstrup,~P.; Popovitz-Biro,~R.; Johnson,~E.; Madsen,~M.~H.; Nyg{\aa}rd,~J.;
  Shtrikman,~H. \emph{Nano Lett.} \textbf{2010}, \emph{10}, 4475\relax
\mciteBstWouldAddEndPuncttrue
\mciteSetBstMidEndSepPunct{\mcitedefaultmidpunct}
{\mcitedefaultendpunct}{\mcitedefaultseppunct}\relax
\EndOfBibitem
\bibitem[Murayama and Nakayama(1994)Murayama, and Nakayama]{Murayama94}
Murayama,~M.; Nakayama,~T. \emph{Phys. Rev. B} \textbf{1994}, \emph{49},
  4710\relax
\mciteBstWouldAddEndPuncttrue
\mciteSetBstMidEndSepPunct{\mcitedefaultmidpunct}
{\mcitedefaultendpunct}{\mcitedefaultseppunct}\relax
\EndOfBibitem
\bibitem[Bao et~al.(2008)Bao, Bell, Capasso, Wagner, M{\aa}rtensson,
  Tr\"{a}g{\aa}rdh, and Samuelson]{Bao08}
Bao,~J.; Bell,~D.~C.; Capasso,~F.; Wagner,~J.~B.; M{\aa}rtensson,~T.;
  Tr\"{a}g{\aa}rdh,~J.; Samuelson,~L. \emph{Nano Lett.} \textbf{2008},
  \emph{8}, 836\relax
\mciteBstWouldAddEndPuncttrue
\mciteSetBstMidEndSepPunct{\mcitedefaultmidpunct}
{\mcitedefaultendpunct}{\mcitedefaultseppunct}\relax
\EndOfBibitem
\bibitem[Spirkoska et~al.(2009)Spirkoska, Arbiol, Gustafsson, Conesa-Boj, Glas,
  Zardo, Heigoldt, Gass, Bleloch, Estrade, Kaniber, Rossler, Peiro, Morante,
  Abstreiter, and an~A. Fontcuberta~i Morral]{Spirkoska09}
Spirkoska,~D. et~al.  \emph{Phys. Rev. B} \textbf{2009}, \emph{80},
  245325\relax
\mciteBstWouldAddEndPuncttrue
\mciteSetBstMidEndSepPunct{\mcitedefaultmidpunct}
{\mcitedefaultendpunct}{\mcitedefaultseppunct}\relax
\EndOfBibitem
\bibitem[Pemasiri et~al.(2009)Pemasiri, Montazeri, Gass, Smith, Jackson,
  Yarrison-Rice, Paiman, Gao, Tan, Jagadish, Zhang, and Zou]{Pemasiri09}
Pemasiri,~K.; Montazeri,~M.; Gass,~R.; Smith,~L.~M.; Jackson,~H.~E.;
  Yarrison-Rice,~J.; Paiman,~S.; Gao,~Q.; Tan,~H.~H.; Jagadish,~C.; Zhang,~X.;
  Zou,~J. \emph{Nano Lett.} \textbf{2009}, \emph{9}, 648\relax
\mciteBstWouldAddEndPuncttrue
\mciteSetBstMidEndSepPunct{\mcitedefaultmidpunct}
{\mcitedefaultendpunct}{\mcitedefaultseppunct}\relax
\EndOfBibitem
\bibitem[Akopian et~al.(2010)Akopian, Patriarche, Liu, Harmand, and
  Zwiller]{Akopian10}
Akopian,~N.; Patriarche,~G.; Liu,~L.; Harmand,~J.~C.; Zwiller,~V. \emph{Nano
  Lett.} \textbf{2010}, \emph{10}, 1198\relax
\mciteBstWouldAddEndPuncttrue
\mciteSetBstMidEndSepPunct{\mcitedefaultmidpunct}
{\mcitedefaultendpunct}{\mcitedefaultseppunct}\relax
\EndOfBibitem
\bibitem[Zhang et~al.(2010)Zhang, Luo, Zunger, Akopian, Zwiller, and
  Harmand]{Zhang10}
Zhang,~L.; Luo,~J.~W.; Zunger,~A.; Akopian,~N.; Zwiller,~V.; Harmand,~J.~C.
  \emph{Nano Lett.} \textbf{2010}, \emph{10}, 4055\relax
\mciteBstWouldAddEndPuncttrue
\mciteSetBstMidEndSepPunct{\mcitedefaultmidpunct}
{\mcitedefaultendpunct}{\mcitedefaultseppunct}\relax
\EndOfBibitem
\bibitem[Heiss et~al.(2011)Heiss, Boj, Ren, Tseng, Gali, Rudolph, Uccelli,
  Peiro, Morante, Schuh, Reiger, Kaxiras, Arbiol, and i~Morral]{Heiss10}
Heiss,~M.; Boj,~S.~C.; Ren,~J.; Tseng,~H.~H.; Gali,~A.; Rudolph,~A.;
  Uccelli,~E.; Peiro,~F.; Morante,~J.~R.; Schuh,~D.; Reiger,~E.; Kaxiras,~E.;
  Arbiol,~J.; i~Morral,~A.~F. \emph{Phys. Rev. B} \textbf{2011}, \emph{83},
  045303\relax
\mciteBstWouldAddEndPuncttrue
\mciteSetBstMidEndSepPunct{\mcitedefaultmidpunct}
{\mcitedefaultendpunct}{\mcitedefaultseppunct}\relax
\EndOfBibitem
\bibitem[Hoang et~al.(2009)Hoang, Moses, Zhou, Dheeraj, Fimland, and
  Weman]{Hoang09}
Hoang,~T.~B.; Moses,~A.~F.; Zhou,~H.~L.; Dheeraj,~D.~L.; Fimland,~B.~O.;
  Weman,~H. \emph{Appl. Phys. Lett.} \textbf{2009}, \emph{94}, 133105\relax
\mciteBstWouldAddEndPuncttrue
\mciteSetBstMidEndSepPunct{\mcitedefaultmidpunct}
{\mcitedefaultendpunct}{\mcitedefaultseppunct}\relax
\EndOfBibitem
\bibitem[Gourley and Biefeld(1984)Gourley, and Biefeld]{Gourley84}
Gourley,~P.~L.; Biefeld,~R.~M. \emph{Appl. Phys. Lett.} \textbf{1984},
  \emph{45}, 749\relax
\mciteBstWouldAddEndPuncttrue
\mciteSetBstMidEndSepPunct{\mcitedefaultmidpunct}
{\mcitedefaultendpunct}{\mcitedefaultseppunct}\relax
\EndOfBibitem
\bibitem[Skowronski and Kremer(1990)Skowronski, and Kremer]{Skowronski90}
Skowronski,~M.; Kremer,~S. T. N. R.~E. \emph{Appl. Phys. Lett.} \textbf{1990},
  \emph{57}, 902\relax
\mciteBstWouldAddEndPuncttrue
\mciteSetBstMidEndSepPunct{\mcitedefaultmidpunct}
{\mcitedefaultendpunct}{\mcitedefaultseppunct}\relax
\EndOfBibitem
\bibitem[not()]{noteNH3}
\emph{All CW and TR PL measurements in the uncapped NWs presented in this work
  were performed after previous NH$_{3}$ treatment for 5 minutes.}\relax
\mciteBstWouldAddEndPunctfalse
\mciteSetBstMidEndSepPunct{\mcitedefaultmidpunct}
{}{\mcitedefaultseppunct}\relax
\EndOfBibitem
\bibitem[Bimberg et~al.(1971)Bimberg, Sondergeld, and Grobe]{Bimberg71}
Bimberg,~D.; Sondergeld,~M.; Grobe,~E. \emph{Phys. Rev. B} \textbf{1971},
  \emph{4}, 3451\relax
\mciteBstWouldAddEndPuncttrue
\mciteSetBstMidEndSepPunct{\mcitedefaultmidpunct}
{\mcitedefaultendpunct}{\mcitedefaultseppunct}\relax
\EndOfBibitem
\bibitem[Titova et~al.(2006)Titova, Hoang, Jackson, Smith, Yarrison-Rice,
  Joyce, Tan, and Jagadish]{Titova06}
Titova,~L.~V.; Hoang,~T.~B.; Jackson,~H.~E.; Smith,~L.~M.;
  Yarrison-Rice,~J.~M.; Joyce,~H.~J.; Tan,~H.~H.; Jagadish,~C. \emph{Appl.
  Phys. Lett.} \textbf{2006}, \emph{89}, 173126\relax
\mciteBstWouldAddEndPuncttrue
\mciteSetBstMidEndSepPunct{\mcitedefaultmidpunct}
{\mcitedefaultendpunct}{\mcitedefaultseppunct}\relax
\EndOfBibitem
\end{mcitethebibliography}

\end{document}